# Emergent rigidity percolation of five-fold aggregates enables controllable glass properties

Wei Chu[1,†], Zheng Wang[1,†], Christopher Ness[2], Konrad Samwer[3,*], Alessio Zaccone[4,3,*], Lina Hu[1,*]

[1] Key Laboratory for Liquid-Solid Structural Evolution and Processing of Materials (Ministry of Education), Shandong University, Jinan, 250061, China
[2] School of Engineering, University of Edinburgh, Edinburgh, EH9 3JL, United Kingdom
[3] I. Physics Institute, University of Göttingen, Friedrich-Hund-Platz 1, 37077 Göttingen, Germany
[4] Department of Physics "A. Pontremoli", University of Milan, 20133 Milan, Italy

[†] These authors contributed equally to this work.
[*] Authors to whom correspondence should be addressed: hulina0614@sdu.edu.cn (L. Hu); alessio.zaccone@unimi.it (A. Zaccone); ksamwer@gwdg.de (K. Samwer)

**Abstract**

Metallic glasses possess outstanding mechanical and physical properties, making them promising candidates for advanced structural and functional applications; however, the lack of understanding and control over their glass transition and solidification processes remains a significant barrier to practical design. The glass transition from liquid to amorphous solid has remained an open problem in physics despite many theories and recent advances in computational efforts. The question of identifying a clear and well-defined diverging length scale accompanying the glass transition has remained unanswered, as has the nature of the transition and, indeed, the presence of a transition at all, as opposed to a mere dynamical crossover. Here we answer these questions using numerical results and theoretical analysis showing that, in atomic (metallic) glass formers, the glass transition coincides with, and is caused by, a continuous rigidity percolation transition from a liquid-like to a solid-like material. The transition occurs as five-fold symmetric atomic clusters progressively aggregate, forming a system-spanning rigid network that marks the onset of mechanical stability. This percolation-driven rigidity growth is accompanied by a sharp increase in the shear modulus $G'$, indicating the emergence of macroscopic solid-like behavior. Beyond this point, which coincides with the Maxwell isostatic point of the percolating structure, dynamical arrest or "freezing-in" prevents further evolution. The long-sought diverging length scale is thus identified as the percolation-driven growth of rigid five-fold clusters, providing a direct link between local structural motifs and macroscopic mechanical properties at the glass transition. These insights offer practical routes to rationally engineer metallic glasses with targeted mechanical stiffness, hardness, and toughness.

# 1. Introduction

Glasses and amorphous solids constitute the largest class of condensed matter on Earth, yet there is no consensus on the true nature of the glass transition. The understanding of amorphous solids and of the glassy state represents one of the biggest open problems in physics [1-5]. The temperature at the glass transition depends on the cooling rate, inconsistent with an equilibrium phase transition scenario. Evidence points to the liquid falling out of thermodynamic equilibrium at the glass transition, which occurs the glass transition temperature denoted $T_g$. Nevertheless, many different theoretical approaches assume that an underlying, "hidden" true or "ideal" phase transition at a temperature above or below $T_g$ may in fact govern the crossover from the liquid to the solid glass [2, 4, 6, 7]. It has also been suggested, however, since Yakov Frenkel, that the glass transition could be a purely kinetic phenomenon or dynamical crossover without any underlying thermodynamic phase transition [4, 8, 9]. Starting from the Adam-Gibbs scenario that postulates cooperatively rearranging regions (CRRs) that grow upon approaching the glass transition, various approaches such as Random First Order Theory (RFOT) have emphasized the role of configurational entropy, associated with the CRRs, which abruptly drops upon approaching $T_g$ [10-13]. This scenario thus predicts the existence of a "diverging" length-scale which has been sought for many decades, although only indicial, and limited, evidence has been found so far, in multi-point correlation functions [14]. Also, these divergencies in multi-point correlation functions do not explain the emergence of rigidity and solid behaviour at the transition, i.e. the emergence of a finite shear modulus ($G'$) at low frequency [15].

A successful theory should explain both the diverging length-scale and the emergence of mechanical stability, as characterized by the development of *G'*, as explained by the nonaffine theory of mechanical response in amorphous solids [16]. So far, no such theory has been presented; the identification of diverging spatial correlations and the analysis of emergent rigidity have remained disjointed.

Metallic glasses, with their disordered atomic networks and metallic bonding, occupy a privileged role within this broad glassy landscape. Its combination of strength, ductility, soft magnetic properties, and biocompatibility has enabled breakthroughs in the development of multilayer composites in areas such as efficient electromagnetic interference shielding [17, 18], to corrosion-resistant Zr-based vascular stents [19], cryogenic-to-high-temperature gears for space actuation [20], and ultra-sensitive torque sensors for robotics [21]. These emerging applications leverage the exceptional soft-magnetic, biocompatible, and mechanical attributes recently reported for bulk metallic glasses [22].

In this paper, we present a unifying scenario based on numerical simulations of a model metallic glass. We demonstrate that the glass transition emerges from a nonequilibrium rigidity-percolation mechanism, providing not only fundamental understanding but also a practical route to precisely control the formation and properties of metallic glasses. The framework is that of an underlying nonequilibrium (continuous or second-order) percolation transition by which atomic clusters with five-fold symmetry aggregate into growing entities which eventually percolate into a system-spanning rigid structure that marks the glass transition. This rigidity percolation transition is nonequilibrium because of the broken detailed balance, which we verify numerically, between aggregation and breakup processes of the growing aggregates, and shares commonalities with the nonequilibrium process governing gelation of attractive colloidal particles [23]. We start with the conceptual theoretical framework, which has much in common with the idea [24] that the glass transition is driven by "the tendency of the atomic species to form locally icosahedral-packed structures in the liquid and for these domains to form extended "polymeric" rigid structures upon approaching the glass transition" recently re-discovered by Douglas and co-workers [25] and ultimately going back to Hägg's original 1935 paper on the "polymerization" mechanism for the glass transition [26]. Here we offer the first quantitative confirmation of this mechanism in a model metallic glass, showing that the rigidity percolation transition is directly reflected in the growth of $G'$. This insight bridges the long-standing gap between microscopic structure and macroscopic mechanical response, thereby enabling rational materials engineering strategies. Because the population of five-fold clusters is known to respond sensitively to minor alloying and cooling rate, our framework naturally translates into qualitative design rules for processing-dependent property control.

## 2. Results

## 2.1 Dynamic Kinetic Model and Theoretical Framework

We consider a dynamic kinetic model where rigid clusters aggregate into larger aggregates, and also break up into smaller ones. The process follows a master kinetic equation [23, 27]:

$$\frac{dc_k}{dt} = \frac{1}{2}\sum_{i,j=1}^{i+j=k} K_{ij}^+ c_i c_j - c_k \sum_{i=1}^{\infty} K_{ik}^+ c_i - K_k^- c_k + \sum_{i=k+1}^{\infty} K_{ik}^- c_i \qquad (1)$$

where $c_i$ is the number concentration of aggregates of size $i$. $K_{ij}^+$ is the rate of association between two aggregates of $i$ and $j$ units, while $K_{ij}^-$ is the breakup rate of an $j+i$ aggregate into smaller ones of size $i$ and $j$, where $i+j=k$. This model is sometimes referred to as a "reversible polymerization" model in the statistical physics literature [27]. The most studied case, with exact solutions, involves i) aggregation taking place between any two aggregates of arbitrary size; ii) breakup involving detachment of dangling units (one-fold coordinated) from an aggregate of arbitrary size. In other words, breakup events leading to two fragments, each larger than one particle, are excluded. This break detailed balance since all aggregation processes are allowed, whereas only breakup processes involving a single detaching unit are allowed [17]. Hence the process proceeds out of chemical and thermodynamic equilibrium [27]. In systems where attraction between units is due to a conservative potential, this assumption is justified. The rationale lies in the high thermal breakup rate of a bond involving an outer-shell dangling unit following Arrhenius' law $\sim \exp[-V/kT]$. This rate surpasses the breakup rate of bonds between units multiply-bonded deep inside the aggregate, which scales at least as $\sim \exp[-nV/kT](n>2)$. Hence in this model [23, 27],

$$K_{ij}^+ = const, \forall i,j$$

$$K_{ij}^- = \lambda K_{ij}^+, if\ i \neq 1, or\ j \neq 1$$

$$K_{ij}^- = 0, if\ i \neq 1, or\ j \neq 1 \qquad (2)$$

Exact solutions for this model [28], describe the cluster size distribution $c_k$. A nonequilibrium continuous phase transition between a sol (i.e., liquid phase) made of clusters, and a gel containing a giant cluster coexisting with small clusters, at $t \to \infty$, where parameter $\lambda$ increases, governing the extent of thermal breakup events where individual units detach from an aggregate. The transition occurs out of equilibrium because detailed balance in the master equation is violated [17]. For $\lambda > 1$, no percolating "gel" forms, whereas a glass transition (or gelation) occurs for $\lambda < 1$, where $\lambda$ is in units of $c_0$ (particles per unit volume). The aggregate size distribution follows a power law in both regimes. At the gel transition, the cluster mass distribution is analytically calculated as $c_k \sim k^{-5/2}$ [23, 27, 28], identifying the Fisher exponent $\tau = 5/2$ for the percolation transition. At the transition, the average aggregate size diverges, and a giant system-spanning percolating aggregate is formed. In standard percolation theory, the power-law form of the cluster mass distribution is a postulate or a guess, whereas here it arises from the analytical solution to the master kinetic equation, Eq. (1), under the assumptions of Eq. (2).

The fractal dimension at the gel point is obtained from the Fisher index $\tau$ using the hyperscaling relation: $\tau = (\frac{d}{d_f}) + 1$. With $d$ =3 and $\tau = 5/2$, $d_f$ is calculated as 2.0 for the clusters, in mean-field theory [17,18]. Within this model, in the liquid phase where only isolated aggregates exist, the cluster mass distribution follows a power-law with an exponential cut-off: $c_k \sim k^{-3/2} \exp(-k/k^*)$, where $k^*$ denotes the upper cut-off in the cluster size. Generally, the rigidity percolation transition is marked by the transition from a cluster mass distribution with an exponential cut-off at a finite mass $k^*$ to a distribution with no exponential cut-off (i.e. with $k^* = \infty$), exhibiting a power-law distribution extending to $k \to \infty$, signalling the emergence of a giant system-spanning percolating aggregate.

While these critical exponents can be obtained analytically under the assumptions of Eq. (2), a similar nonequilibrium percolation due to irreversible cluster growth can occur with different aggregation and breakup rates. This implies that such transition, possibly with different exponents and a different universality class, may still occur even if breakup of clusters does not proceed exactly by removal of one unit at the time. For example, when two units break up from a cluster instead of one, the asymmetry between the aggregation mechanism (involving all cluster sizes) and the breakup mechanism (involving a smaller subset of sizes) will still generate a nonequilibrium percolation transition with similar features but possibly different critical exponents.

## 2.2 Model system and selection of characteristic atoms

Following recent work highlighting the existence of growing five-fold symmetry clusters upon approaching the glass transition [29], our study quantitatively analyzes if these aggregates facilitate a nonequilibrium percolation transition along the lines proposed by Hägg [27] and Ref. [24]. Therefore, we performed molecular dynamics (MD) simulations of the $Cu_{50}Zr_{50}$ alloy using LAMMPS software with embedded atom potentials [30-34]. Specific details of the modelling are given in the Experimental Section. Figure 1(a) illustrates the temperature dependence of the quantity $E$-$3k_BT$ obtained from our simulations, clearly identifying the glass transition temperature $T_g$ by a distinct crossover in slope. Figure 1(b) shows the interatomic potential employed in our simulations, highlighting that the first intersection point of the potential was selected as the cutoff distance for defining atomic clusters. Specific details of the simulation protocols are provided in the Experimental Section, while sensitivity tests of model parameters, justification for choosing this cutoff distance, and additional validation details are available in the Supplementary Materials (Figures S1–S6). We note that similar five-fold populations have been reported in Cu–Zr–(Al) and Fe-based glasses, indicating that the mechanisms discussed below are not limited to the present alloy family.

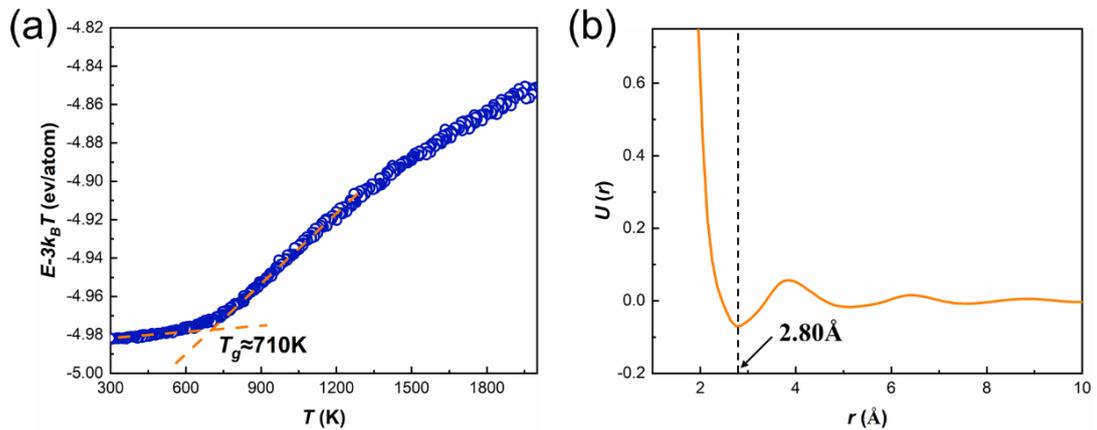

**Figure 1.** Simulation details of the metallic liquid. (**a**) Temperature dependence of $E-3k_BT$ for $Cu_{50}Zr_{50}$ liquid upon cooling. The crossover temperature is confirmed as the glass transition temperature $T_g$. (**b**) The potential function used in this work. The first intersection of the potential function is defined as the cutoff distance for whether different atoms are considered to be in the same cluster.

To explore the onset of percolation of the five-fold clusters and its relation to the glass transition, we focus on the evolution of the average coordination number, $Z$. The coordination number $Z$ is defined as $(1/N)\sum_{i=1}^{N} N_{b,i}$, where $N_{b,i}$ represents the number of bonds of atom $i$, and $N$ is the number of atoms. This study primarily considers atoms with local five-fold symmetry (LFFS, $f^5$) greater than 0.6, a threshold established in previous work [35, 36]. Specifically, Hu et al. [36] found that these atoms tend to aggregate in metallic liquids. Due to high five-fold symmetry, they also correspond to icosahedral-like structures. Icosahedra where an atom at the center is bonded to 12 neighbors are the hallmark of thermodynamically stable amorphous liquid-like structures [36]. Such icosahedral clusters significantly proliferate upon cooling and strongly influence the properties of the resultant amorphous solids [37, 38].

According to group theory, lattices with five-fold symmetry cannot support long-range translational order [39]. Except for exotic cases like quasicrystals, five-fold clusters are inherently incapable of forming dense, periodically repeating lattices. Instead, they randomly aggregate into progressively larger clusters, analogous to colloidal particles dispersed in a solvent. Upon cooling, icosahedra, favored for their high atomic packing density, are promoted by the cooling process, which densifies the system. This leads to the growth of aggregates of five-fold clusters with an effective kinetic rate denoted by $K_{ik}^+$. Although the exact nature of $K_{ik}^+$ is not specified, five-fold clusters are expected to merge due to both densification-driven compression and internal elastic stresses arising from local geometric frustration inherent in five-fold orientational symmetry. Besides densification, the merging of five-fold clusters may also reduce local elastic energy associated with such geometrical frustration. Interestingly, at $T_g$, no abrupt volume change is observed, instead, a clear slope change in the thermal expansion coefficient (by approximately a factor of 3) is detected [40]. We also performed quantitative analyses to investigate the effect of other coordination symmetries. Figure 2(a) and (b) show the distributions of the four-fold and six-fold symmetries at different temperatures. These distributions illustrate that four-fold and six-fold symmetries are predominantly present at lower symmetry values (below 0.5), in sharp contrast with the high symmetry region occupied by five-fold clusters. This difference underscores the uniqueness of five-fold symmetry in forming larger and more stable clusters characteristic of the amorphous phase.

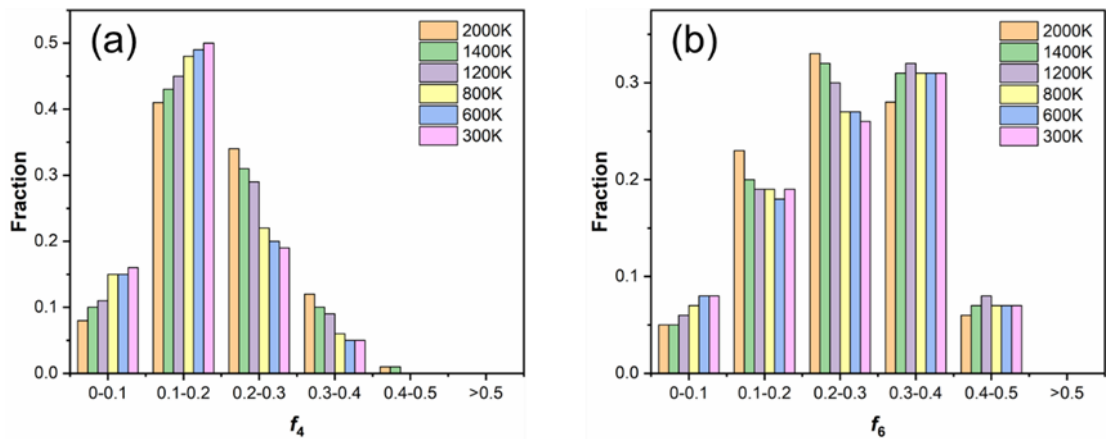

**Figure 2.** The distributions of (**a**) four-fold, and (**b**) six-fold symmetries at six different temperatures. The x-axis indicates the range of local symmetry with different degrees, e.g., 0.1-0.2 in Figure 2(**a**) indicates that the fraction of four-fold symmetry is greater than 0.1 but less than or equal to 0.2.

## 2.3 Percolation Dynamics and Evolution of Clusters

Due to their intrinsic stability, the five-fold atomic clusters exhibit significantly lower decay rates compared to the system average, reflecting their inherently slower atomic mobility. Consequently, the fraction of atoms in the largest aggregate formed by five-fold stable clusters grows upon cooling towards the glass transition. We plot the fraction of atoms in the largest cluster, $f_z$, as a function of $Z$ in Figure 3(a). We observe a sharp increase in $f_z$ as the system approaches the critical coordination number $Z_c \approx 6$, indicating rapid growth of the largest five-fold aggregate. Notably, we identify a clear crossover in the scaling exponent $\sigma$, defined by the relation $f_z \sim (Z_c - Z)^{-\sigma}$, occurring near this critical value $Z_c$, which corresponds precisely to the glass transition temperature $T_g$. The critical value ($Z_c \approx 6$) was also rigorously determined by continuously cooling close to 0 K to find its maximum value (see Figure S7, Supporting Information), essentially matching our expectation of 6. Importantly, this critical value of connectivity coincides with the isostatic Maxwell rigidity condition, at which a finite value of the low-frequency shear modulus $G'$ sets in, according to $G' \sim (Z - 6)$, as predicted by nonaffine response theory of spherical particles interacting via non-covalent potentials [41, 42]. Figure 3(b) shows that $Z$ monotonically increases with decreasing temperature, approaching the unique saturation value $Z = 6$ at low temperatures. This behavior closely parallels that observed in nonequilibrium percolation (gelation) transitions of attractive colloidal systems, extensively studied experimentally and through simulations [23], thereby strongly supporting our proposed scenario. The present findings clearly indicate that the nonequilibrium percolation transition precisely coincides with the onset of rigidity and the liquid-to-solid transition - hallmarks of the glass transition.

To gain deeper insights into the percolation dynamics, we further analyze the fraction of particles participating in five-fold clusters as a primary structural indicator. As depicted in Figure 3(c), this fraction increases monotonically upon cooling, exhibiting a pronounced rise around the glass transition temperature. This quantity directly measures atomic participation in the percolation process, explicitly showing how characteristic structural motifs drive connectivity evolution during cooling. Additionally, Figure 3(d) quantifies the "strength" of percolating clusters, by showing the fraction of atoms belonging to system-spanning clusters. In the absence of system-spanning clusters, the largest available cluster is used for analysis. This approach provides essential insights into the development of large-scale connectivity as the system approaches percolation conditions. Notably, the clear emergence of spanning clusters as the temperature nears $T_\mathrm{g}$ conclusively demonstrates the coincidence between the glass transition and the percolation threshold, highlighting the intimate relationship between structural connectivity and the onset of mechanical rigidity.

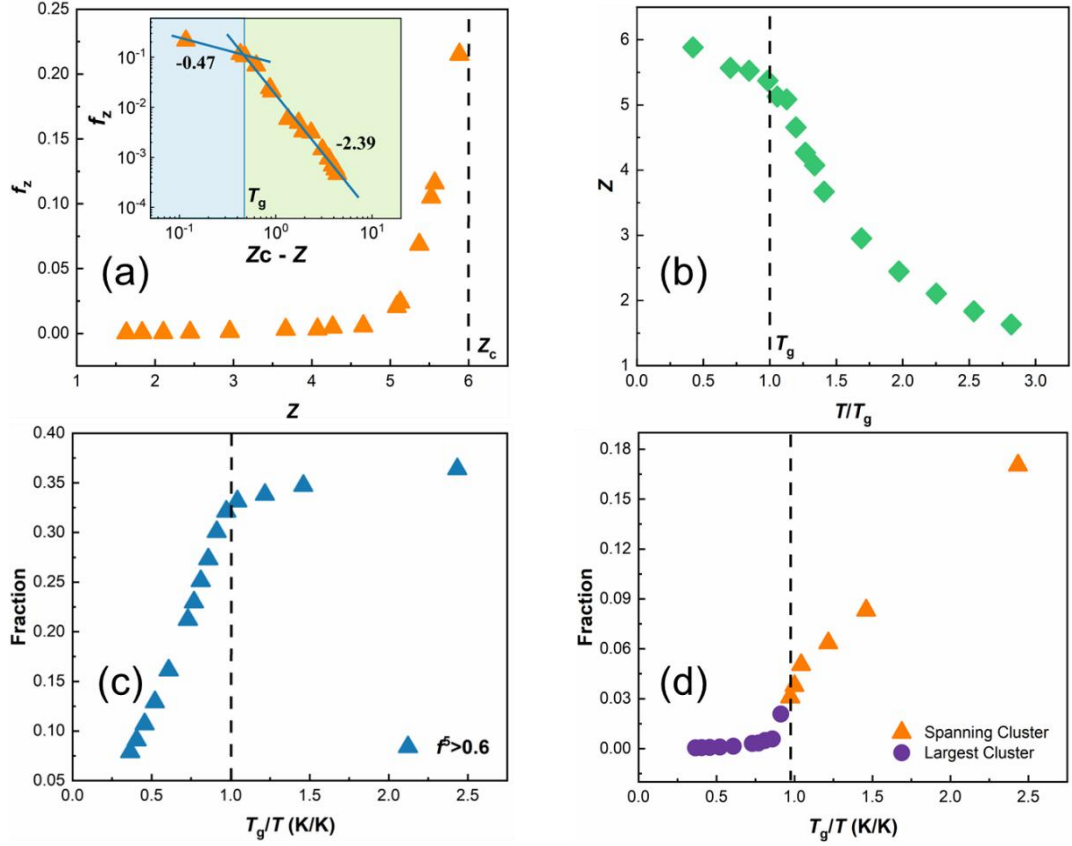

**Figure 3.** Evolution of cluster size during percolation and glass transition processes, based on molecular dynamics simulations. (**a**) Fraction of atoms in the largest cluster, $f_z$, as a function of the average coordination number $Z$. Inset: Evolution of $f_z$ upon approaching the glass transition temperature $T_g$. (**b**) Average coordination number $Z$ as a function of temperature in the simulation. (**c**) Fraction of particles that participate in the percolation process. (**d**) Fraction of particles in the spanning clusters. When there are no spanning clusters in the system, the largest cluster is used instead.

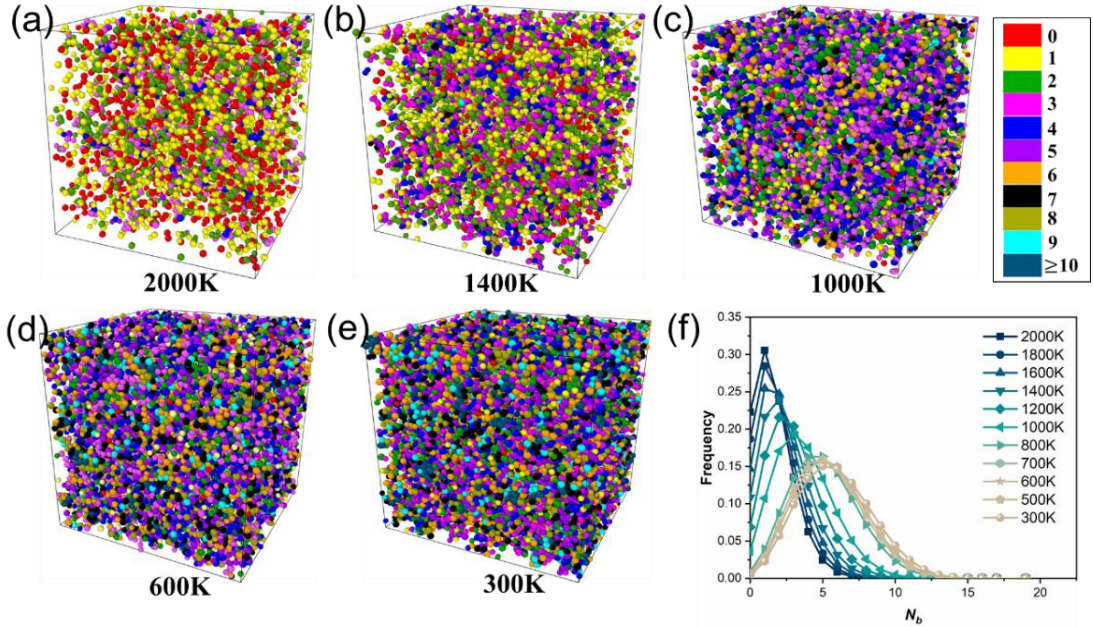

**Figure 4.** Observation of "gelation" in simulations. (**a**)-(**e**) Reconstructions and (**f**) Bond histograms of metallic atoms with $f^5 > 0.6$ at different temperatures in simulation.

To visually elucidate how the nonequilibrium percolation transition develops at the atomic scale, we reconstructed the spatial distribution of atoms with five-fold symmetry (LFFS, $f^5 > 0.6$) at selected temperatures in Figure 4(a-e). At higher temperatures (1600 K, 1400 K), clusters are mostly small, isolated entities dispersed throughout the liquid, indicative of a dilute clustering regime dominated by frequent atomic rearrangements. Upon further cooling (1200 K, 900 K), the clusters progressively grow larger, and local connectivity markedly increases, signaling enhanced local structural ordering driven by thermally activated aggregation. Approaching the glass transition temperature (700 K), a system-spanning cluster emerges, revealing a clear structural signature of percolation. The accompanying bond distribution histograms (Figure 4(f)) quantitatively confirm this progressive change in atomic environments. With decreasing temperature, the histogram distinctly shifts toward higher coordination numbers, clearly demonstrating the formation of increasingly well-connected structures. At high temperatures, bonds are predominantly few, corresponding to loosely connected or isolated clusters, whereas near $T_g$, the bond distributions peak at significantly higher values, highlighting the emergence of extensive, rigid network structures characteristic of the solid-like amorphous state. These microscopic visualizations and quantitative distributions thus explicitly capture the evolution from isolated atomic clusters toward a densely interconnected, percolating network underpinning the rigidity onset at the glass transition.

## 2.4 Structural Analysis of Growing Clusters

The correlation length of connected atoms was calculated using $\xi^2 = 2\sum_i R_{gi}^2 N_i^2 / \sum_i N_i^2$, where $R_{gi}$ is the radius of gyration for clusters of size $N_i$ [43]. Figure 5(a) illustrates how $\xi$ increases with the average coordination number $Z$ during cooling. As the system approaches the glass-transition regime, the correlation length $\xi$ diverges, coincides with the nonequilibrium percolation of five-fold clusters. The inset of Figure 5(a) explicitly tracks the scaling relationship $\xi \sim (Z_c - Z)$, early indicating an initial gradual upturn near the onset temperature $T_c$, followed by a much sharper increase upon nearing the glass transition temperature $T_g$. This behavior explicitly demonstrates that the percolation process initiates around $T_c$ and culminates at $T_g$.

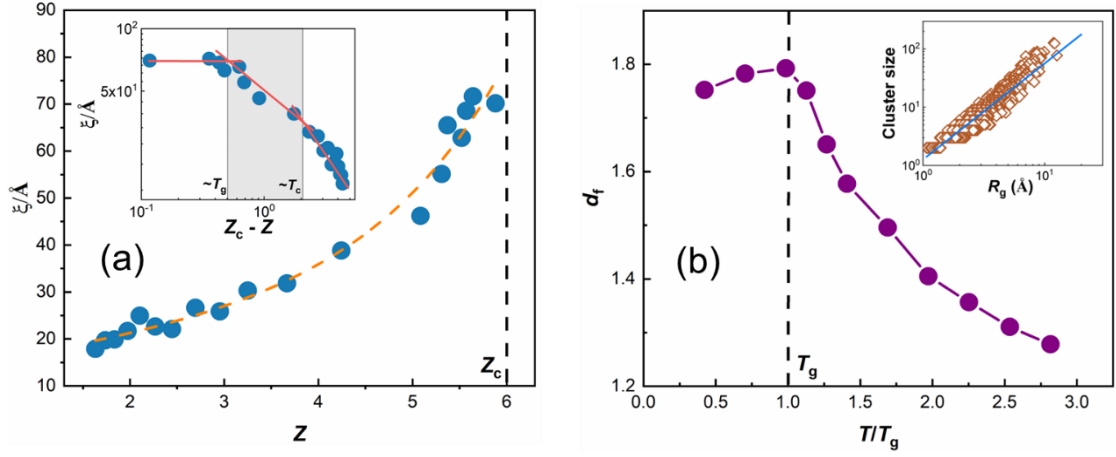

**Figure 5.** (**a**) Correlation length, $\xi$, as a function of the average coordination number $Z$. Inset: Evolution of $\xi$ during the cooling process. (**b**) Fractal dimension $d_f$ as a function of temperature in simulations. Inset: Cluster size versus radius of gyration at 900 K, as an example.

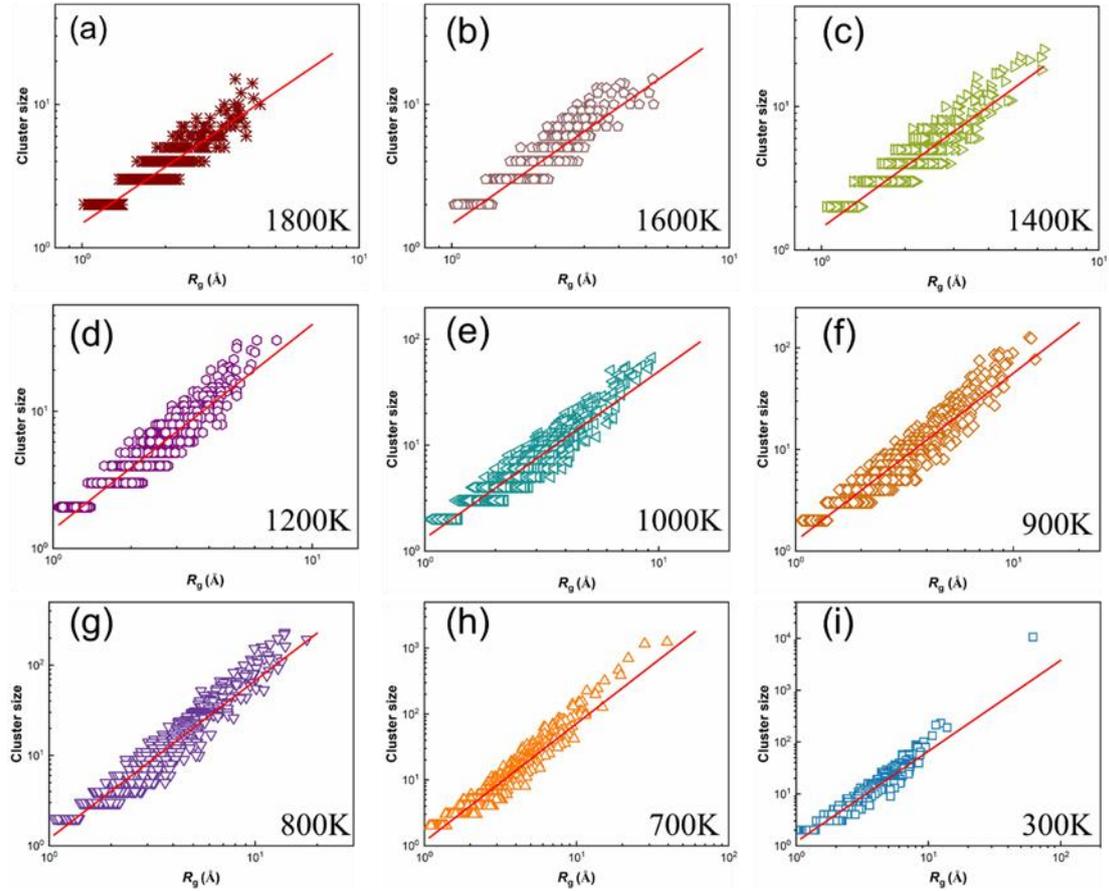

**Figure 6.** Cluster size versus radius of gyration in simulations at different temperature.

Each point represents a cluster.

By employing the hyperscaling relation of critical phenomena, we theoretically expect the fractal dimension $d_f = 2.0$ for the percolating clusters, based on Eq. (1) [13]. To test this prediction, we determined the fractal dimension from simulations by plotting cluster size against the radius of gyration ($R_g$), as shown in Figure 5(b). Figure 6 further presents detailed data showing cluster size versus radius of gyration at multiple temperatures, where each data point represents an individual cluster. As the temperature decreases, $d_f$ increases monotonically and then remains at a relatively stable value (~ 1.80) once the percolation transition is completed at temperatures around $T_g$. An increase in the fractal dimension leads to a more compact structure [44, 45]. It is noteworthy that the observed fractal dimension ($d_f$ ~ 1.80) deviates slightly from the ideal theoretical value of 2.0. This deviation can be rationalized because we focus exclusively on five-fold clusters as the characteristic structural group; hence, the critical exponents need not exactly match those previously reported for colloidal gelation [13]. Moreover, deviations from the mean-field prediction ($d_f = 2.0$) may arise from intrinsic non-mean-field effects, including finite-size effects, short-range interactions, or spatial heterogeneity in the metallic glass-forming system studied here.

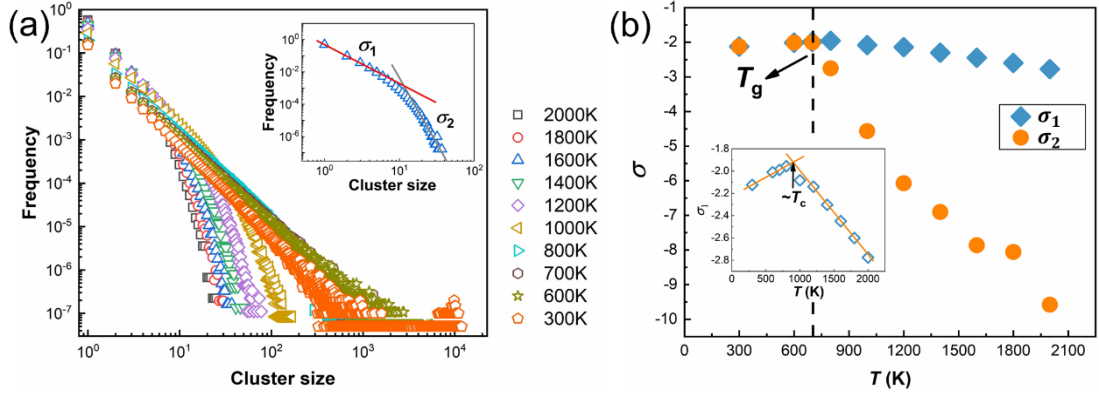

**Figure 7.** Cluster-size distribution across the percolation transition, derived from molecular dynamic simulations. (**a**) Cluster-mass distributions during the cooling process. Inset: A power-law fit of the cluster-mass distribution (example at 1600 K). (**b**) The exponent $\sigma$ of the power-law fit (frequency ~ (cluster size)) to the cluster size distribution, presented as a function of temperature. Inset: The exponent $\sigma_1$ of the power-law fit.

## 2.5 Cluster-Mass Distribution and Critical Exponents

Figures 7(a) and (b) present the evolution of cluster sizes during the cooling process. As the temperature decreases, the cluster-size distribution gradually shifts to larger sizes, i.e. to larger cut-off values $k^*$, and the power-law distribution emerges. Eventually, around 800 K, the exponential cut-off is no longer present, meaning that $k^* \to \infty$, and the cluster mass distribution follows a power-law over the entire range of accessible sizes. This signals the divergence of the cluster mass distribution at the nonequilibrium percolation transition. The evolution of the power-law exponent $\sigma$ with temperature is shown in Figure 7(b). In colloidal systems, it has been reported that $\sigma$ typically decreases from -3/2 to -5/2 across the gelation transition, in agreement with mean-field theoretical predictions [23]. However, for the metallic liquid studied here, the cluster-mass distributions at higher temperatures cannot be fully captured by a single power-law (inset of Figure 7(a)). Instead, two distinct power-law exponents, denoted as $\sigma_1$ and $\sigma_2$, for small and large cluster sizes respectively, were extracted. Both exponents $\sigma_1$ and $\sigma_2$ increase monotonically and then stabilize as the system approaches the glass transition temperature $T_g$, as shown in Figure 7(b). This gradual evolution of power-law exponents can be rationalized by considering the theoretical relationship $c_k \sim k^{-3/2} \exp(-k/k^*)$, at higher temperatures, the presence of a smaller size cut-off $k^*$ naturally results in smaller exponent values. Such a gradual change in exponent values is notably absent in colloidal gels during their sol phase but is distinctly characteristic of rigidity percolation in metallic glasses, marked by the change in $\sigma$ (Figure 7(b)). After crossing the percolation threshold, the critical value $\sigma$ of stabilizes around -2, somewhat larger than the mean-field theoretical prediction of -5/2. his deviation likely originates from the inherent assumptions of mean-field theory or from structural factors specific to the metallic glass system studied here, such as a comparatively less compact structure of atomic clusters. Furthermore, a closer inspection of the temperature dependence of $\sigma_1$ (inset of Figure 7(b)), clearly indicates that the onset of percolation is detectable near the characteristic temperature $T_c$. Importantly, excluding system-spanning clusters from statistical analysis was found not to qualitatively alter the observed trends or conclusions, thereby confirming the robustness of our findings regarding the percolation transition.

## 2.6 Violation of Detailed Balance and Non-equilibrium Dynamics

To verify that the observed percolation transition of five-fold clusters (which coincides with the glass transition) represents a nonequilibrium percolation process similar to that observed in colloidal systems [23], we analyzed the aggregation and breakup kinetics of five-fold clusters. Specifically, we measured the aggregation rate $K_{ik}^{+}$, and the breakup rate $K_{ij}^{-}$ for the aggregation and breakup rate of single bonded clusters to and from an aggregate, respectively, upon varying the aggregate size, as shown in Figure 8. Across all temperatures investigated, aggregation rates consistently exceeded breakup rates, and the gap between the two rates widened significantly upon cooling. Notably, at approximately 700 K (close to the glass transition temperature $T_g$), the aggregation rates became independent of cluster size, implying that clusters of any size aggregated at similar rates, a scenario well-aligned with the predictions from Eq. (2). In contrast, the breakup rates showed a pronounced dependence on cluster size, rapidly decreasing as the size of clusters increased, with a sharp peak associated specifically with the breakup of single-particle units. This marked asymmetry between the aggregation and breakup explicitly demonstrates a violation of detailed balance, as described by the master kinetic equation, Eq. (1) [17]. Although the exact assumptions of Eq. (2) are not strictly satisfied here - since breakup events in our simulations are not exclusively limited to single-particle detachment - the core characteristics of a nonequilibrium percolation transition remain clearly evident. These findings robustly support our interpretation that the glass transition in metallic liquids is governed by a nonequilibrium percolation mechanism, consistently matching the behavior of the cluster-mass distribution exponents previously presented in Figure 7(b).

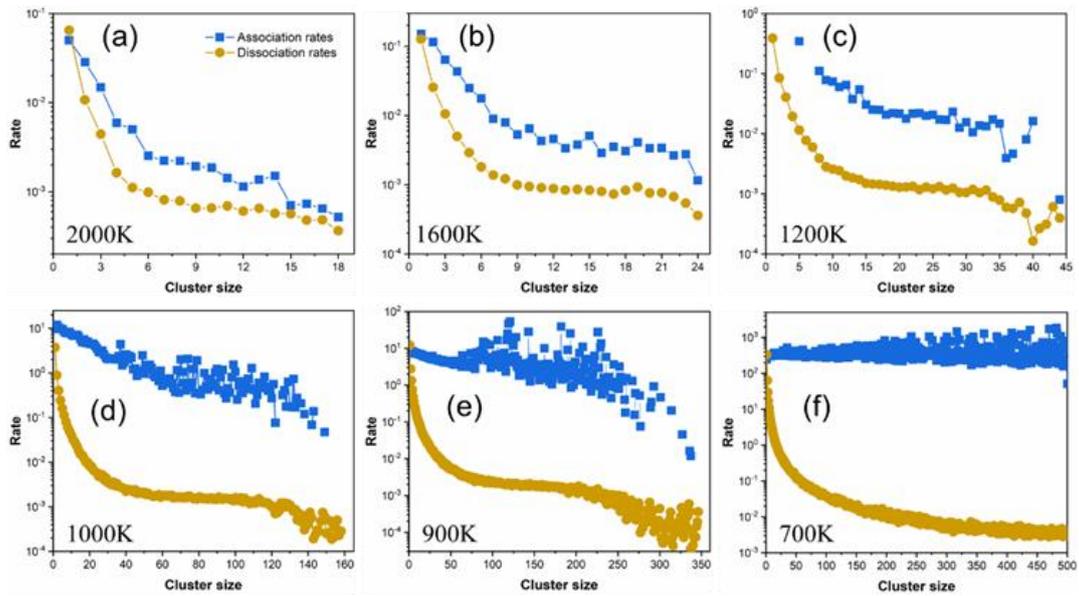

**Figure 8.** Violation of detailed balance through molecular dynamics simulations in metallic system ($Cu_{50}Zr_{50}$). Aggregation rates (blue symbols) and breakup rates (yellow symbols) of single-bonded atomic clusters of different cluster sizes in simulation at different temperatures. The results demonstrate that aggregation becomes cluster-size independent near the glass transition temperature $T_g$, while breakup rates sharply decrease with cluster size, highlighting the asymmetry between aggregation and breakup processes.

## 2.7 Cluster Lifetime and Structural Relaxation Dynamics

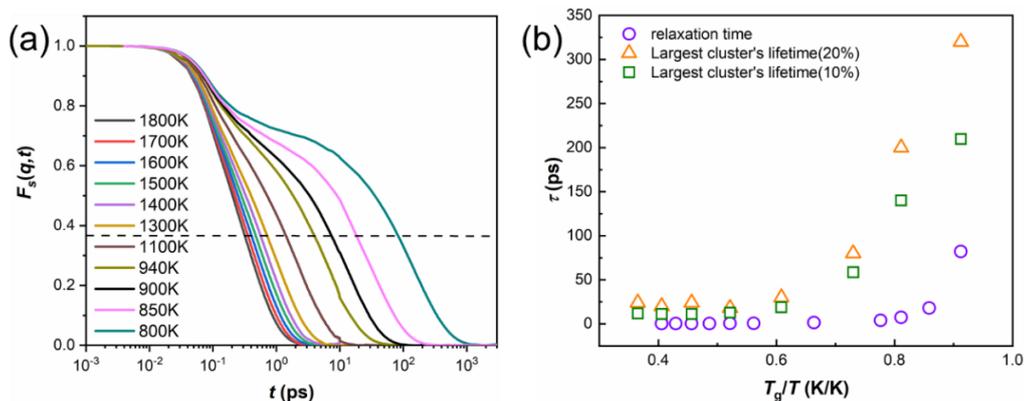

**Figure 9.** (**a**) Self-intermediate scattering functions (SISFs) of all atoms at different temperatures. Here, the structural relaxation time was determined by the time at which the SISFs decays to e$^{-1}$ (dash line). (**b**) Comparison between the structural relaxation time with the "lifetime" of the largest cluster.

To further clarify the dynamical correlation between cluster percolation and structural relaxation during the glass transition, we analyzed the lifetime of clusters and directly compared it with the structural relaxation time of the system. The key scientific motivation behind this analysis is to explicitly verify whether structural arrest at the glass transition can be directly linked to the persistence of dynamically stable atomic clusters, providing quantitative support for our hypothesis of a nonequilibrium percolation-driven glass transition. The structural relaxation dynamics were characterized using the self-intermediate scattering function (SISF) [46], defined as:

$$F_S(q,t) = \frac{1}{N} \langle \sum_{j=1}^{N} e^{iq \cdot [r_j(t) - r_j(0)]} \rangle \tag{3}$$

where $r_j(t)$ is the position of atom *j* at time *t*, and *q* is typically chosen to correspond to the first peak of the static structure factor. Figure 9(a) shows SISFs at various temperatures, clearly illustrating that structural relaxation slows significantly upon cooling. We defined the structural relaxation time, $\tau_\alpha$, as the time at which $F_S(q,t)$ decays to *e*$^{-1}$, as indicated by the dashed line in Figure 9(a). In parallel, we examined the dynamical stability of atomic clusters by monitoring the lifetime of the largest cluster. The cluster lifetime is defined practically as the duration from when the largest cluster first forms until its size subsequently reduces by more than 10%. As depicted in Figure 9(b), both the structural relaxation time $\tau\alpha$ and the largest cluster lifetime increase markedly as temperature decreases, closely mirroring each other's trends. This clear correlation demonstrates that the slowing down of structural relaxation - a hallmark of glass formation - is intimately connected to the lifetime of system-spanning clusters. Such dynamical correspondence provides strong evidence for the crucial role of cluster percolation in governing the structural arrest observed at the glass transition.

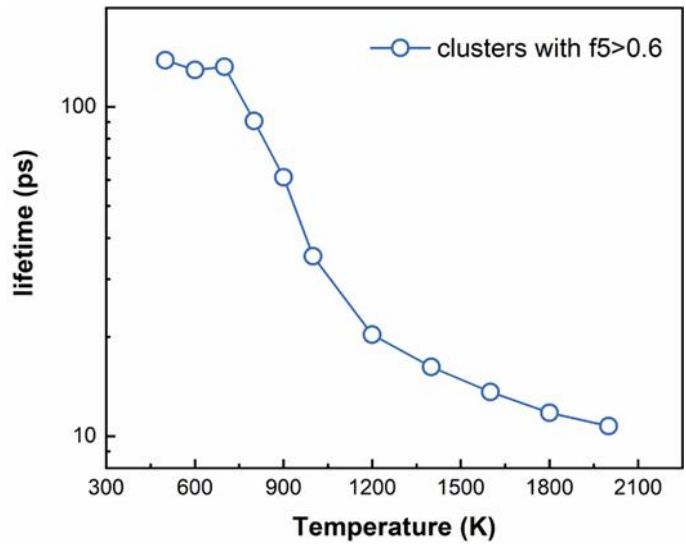

**Figure 10.** The lifetime of clusters with $f^5 > 0.6$ at different temperature.

Additionally, we investigated the lifetime of stable five-fold symmetry clusters (defined by $f^5 > 0.6$) explicitly at different temperatures, as presented in Figure 10. The lifetime of these stable five-fold clusters monotonically increases as the system cools, exhibiting a pronounced enhancement as the glass transition temperature $T_g$ is approached. Notably, the lifetime of these characteristic clusters tends to saturate at temperatures below $T_g$, clearly signifying that the clusters have reached a long-lived or effectively permanent configuration characteristic of the amorphous solid state. The central significance of these findings lies in confirming that stable, five-fold symmetric clusters serve as robust structural motifs that kinetically trap the system in an amorphous state, thereby underpinning the long-term mechanical stability associated with metallic glasses.

## 2.8 Viscoelastic Response and Emergence of Rigidity

We further investigated the viscoelastic response of samples of different sizes under oscillatory shear using molecular dynamics-based dynamic mechanical spectroscopy (MD-DMS). A typical oscillatory shear measurement is shown in Figure 11(a), where the temperature is 300 K and the oscillation period is $t_p$ = 100 ps. In these measurements, the fitted phase shift $\delta$ and $\sigma_A$, were used to calculate the storage modulus $G' = \sigma_A/\varepsilon_A * \cos(\delta)$ and the loss modulus $G'' = \sigma_A/\varepsilon_A * \sin(\delta)$, respectively. The storage and loss moduli at different temperatures were obtained by performing MD-DMS simulations with various oscillation periods. The modulus data were from eight different samples, including three large systems (L-50, L-100, and L-1000) with 54,000 atoms each, and five small systems (S-100, S-200, S-500, S-2000, and S-4000) with 16,000 atoms each. Figures 11(b) and (c) present the temperature dependence of the storage modulus ($G'$) and loss modulus ($G''$). In all samples, a pronounced transition is observed near $T_g$, the storage modulus $G'$ exhibits a sharp increase, marking the onset of mechanical rigidity, while the loss modulus $G''$ decreases correspondingly. In addition to mechanical responses, Figure 11(d) examines the structural evolution by tracking the fraction of five-fold atoms, as well as the size of the largest cluster (LC) and the second-largest cluster (SLC) among these five-fold clusters, as a function of temperature. At high temperatures, LC and SLC are comparable in size, indicating the absence of long-range connectivity. As the temperature approaches $T_g$, the largest cluster (LC) rapidly grows and separates from the SLC, signaling the onset of a percolating rigid network. This structural change coincides with the sharp rise in $G'$, establishing a direct correlation between cluster aggregation and mechanical stiffening. These results demonstrate that the aggregation of five-fold symmetric clusters under cooling drives the formation of a system-spanning network, which underlies the macroscopic glass transition and the emergence of mechanical rigidity.

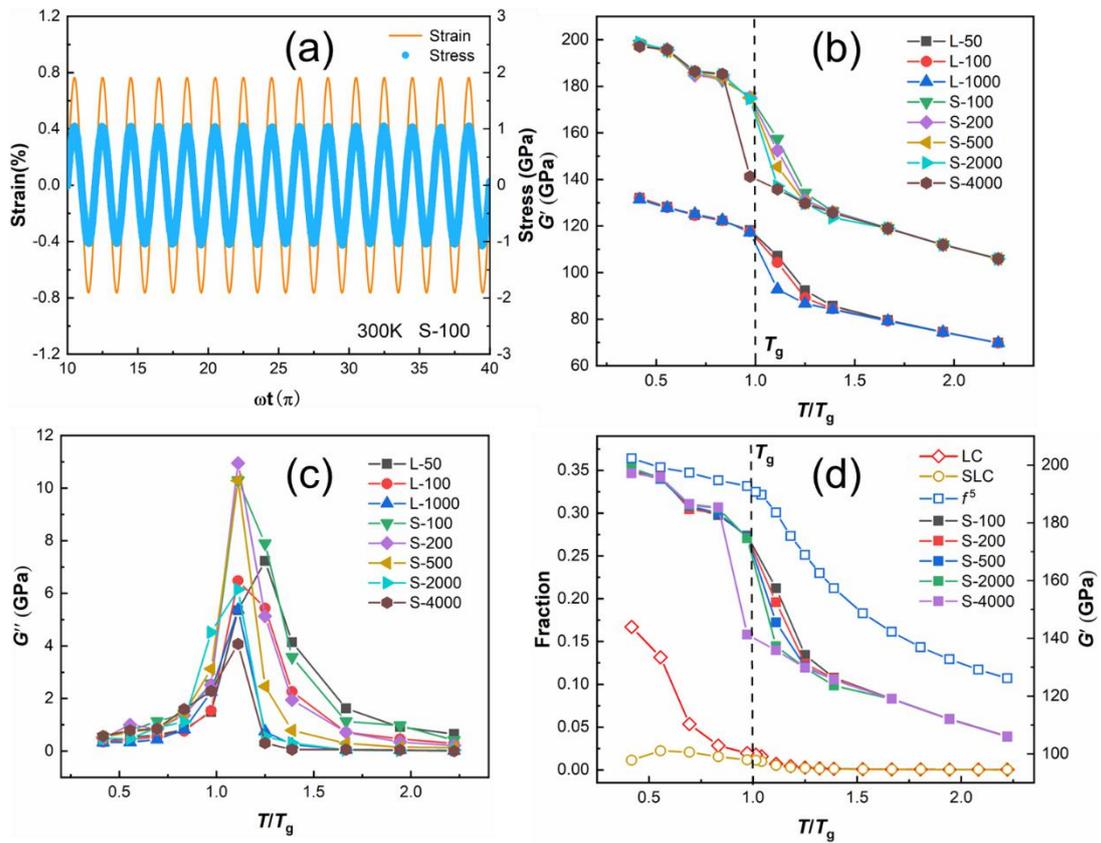

**Figure 11.** Temperature dependence of the viscoelastic response and structural evolution in samples with different sizes, highlighting the correlation between five-fold

cluster aggregation and the emergence of mechanical rigidity. (**a**) A typical MD-DMS measurement at $T = 300$ K for the S-100 sample, showing the applied sinusoidal strain (left axis) and the resulting stress response (right axis). (**b**) Storage modulus $G'$, and (**c**) loss modulus $G''$, as functions of temperature $T/T_g$, for various samples. (**d**) Fraction of five-fold symmetric clusters, and the size evolution of the LC, SLC in five-fold clusters, along with the storage modulus $G'$, plotted as functions of $T/T_g$, respectively. The vertical dashed line marks the glass transition temperature $T_g$.

We also provide direct visualizations of the nonequilibrium percolation process associated with the glass transition, as presented in Figure 12. Molecular dynamics simulations reveal that as the system is cooled from high temperatures, atomic clusters gradually grow in size. Initially, at high temperatures, the clusters are small and dispersed. As the temperature approaches the onset temperature $T_c$, the ten largest clusters (highlighted in yellow) begin to expand significantly and start to interconnect, signaling the initiation of rigidity percolation. Around $T_c$, isolated clusters merge and a system-spanning percolating network (highlighted in red) emerges. Upon further cooling below the glass transition temperature $T_g$, this percolated structure becomes fully established, with the largest clusters extending continuously across the entire simulation domain. This indicates that a nonequilibrium "gelation"-like transition occurs, which coincides with the mechanical rigidity and arrest of atomic motion characteristic of the glass transition. The sequential aggregation of atoms at different temperatures provides a clear visual demonstration of the percolation-driven mechanism underlying glass formation.

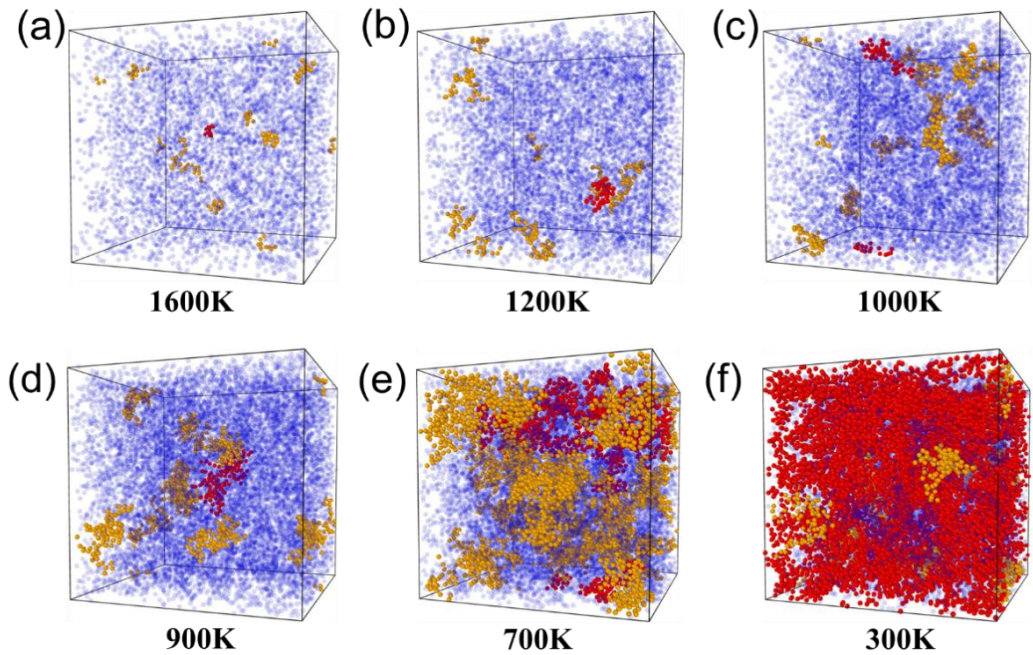

**Figure 12.** Observation of nonequilibrium percolation ("gelation") transition coinciding with the glass transition, based on molecular dynamics simulations. Observation of aggregating metallic atoms at different temperatures. The largest ten atom clusters (yellow) were selected to observe the aggregation of atoms at different temperatures, and the largest clusters are marked in red.

## 2.9 Implications for Plastic Deformation and Toughness

Our findings establish that the emergence of macroscopic rigidity in metallic glasses originates from the percolation of stable five-fold symmetric atomic clusters. This percolating network underpins the rise of the shear modulus $G'$, signaling the transition from liquid-like to solid-like mechanical behavior. However, while this connectivity is crucial for achieving high stiffness and strength, its inherent spatial heterogeneity also has important consequences for plastic deformation. In particular, regions lacking sufficient five-fold connectivity may act as local soft spots, facilitating the nucleation of shear transformation zones (STZs) [47-49]. These localized zones can concentrate strain and initiate shear bands, leading to premature failure. From a structural-design perspective, this dual role of connectivity implies that the topology and distribution of the five-fold network not only control the onset of rigidity, but also influence the material's resistance to shear localization. Engineering a more uniform or fragmented percolating network, through alloying strategies or controlled quenching, could promote the multiplication and arrest of shear bands, thereby improving ductility without compromising strength. This perspective links the microscopic rigidity landscape revealed in our work to the macroscopic toughness challenge in metallic glasses, and motivates future efforts to couple percolation-based structural metrics with deformation mapping under load. Such understanding may provide new pathways for designing metallic glasses with both high strength and enhanced plastic compliance. Moreover, by grounding plasticity and toughness in the same percolation framework that governs the glass transition itself, these results suggest a unifying structural metric for amorphous solids, metallic or otherwise, linking the emergence of rigidity, the localization of flow defects, and the ultimate mechanical limits of glasses.

## 3. Conclusions

In summary, we demonstrated that the glass transition in supercooled liquids coincides with, and is caused by, a nonequilibrium rigidity percolation transition of stable five-fold atomic clusters, coinciding with the mechanical rigidity transition at the Maxwell isostatic point. This point marks where negative nonaffine contributions to the shear modulus equal the positive affine contributions, and the shear modulus is vanishing upon heating. Molecular simulations provided supporting evidence by showing the diverging length scale at the glass transition, given by the correlation length of the rigid percolating cluster. The nonequilibrium nature of the percolation transition was confirmed by the asymmetry between aggregation and breakup rates, violating the detailed balance in the master kinetic equation. Overall, our results provide the long-sought answer to the nature of the glass transition, explaining it as an underlying nonequilibrium continuous rigidity percolation transition of five-fold atomic clusters. This mechanism also explains the liquid-to-solid transition as the formation of a rigid system-spanning amorphous network of stable five-fold clusters, driven by densification and local frustration, which leads to the emergence of macroscopic rigidity, as evidenced by the sharp increase in shear modulus $G'$ near the glass transition. The transition shares key features with rigidity-percolation phenomena in other disordered media, where critical clusters display conformal invariance [50] and fractal growth governed by universal exponents [51], indicating that metallic-glass formation belongs to a broader universality class linking structural disorder to emergent elasticity. Because the connectivity of the five-fold network can be tuned through alloy composition and cooling protocol, our findings also provide a tangible route to tailor macroscopic properties - such as modulus, damping, and toughness - in next-generation metallic-glass components.

## 4. Experimental Section

*Molecular dynamics (MD) simulations:* Molecular dynamics simulations were performed to model trajectories of atoms of $Cu_{50}Zr_{50}$ using the LAMMPS software [52]. The Embedded Atom Method potential developed by Sheng et al. is used to describe the atomic interaction between Cu and Zr atoms [31]. Simulations are performed in a cubic box with periodic boundary conditions containing $N = 54000$ atoms. The samples are first melted at 2000 K and relaxed for 2 ns under the NPT ensemble (constant number of atoms, pressure, and temperature) to get an equilibrium liquid. Subsequently, the systems were cooled from 2000 K down to 300 K at a controlled cooling rate of $10^{12}$ K/s. Configurations at different intermediate temperatures were systematically collected for subsequent calculations. At each temperature, the configurations were allowed to stabilize under controlled conditions, ensuring that macroscopic properties reached a steady state suitable for data collection. After stabilization, 1000 configurations were systematically collected at regular intervals of 40 fs for detailed subsequent analysis. The time step was set to 2 fs and the temperature and pressure were controlled using a Nose-Hoover thermostat and a Nose-Hoover barostat, respectively [32]. Voronoi tessellation analysis and the Open Visualization Ovito software were used to characterize the structure [33]. It is important to note that achieving true thermodynamic equilibrium near or below the glass-transition temperature ($T_g$) is challenging due to the extremely long relaxation times associated with deeply supercooled liquids and glasses. Therefore, although our simulations aim to capture the essential features of the system's behavior at each temperature studied, achieving true equilibrium in simulations at and below $T_g$ is practically challenging, if not impossible. However, the approach we take is quite general, which ensures the validity of our results. This method ensures that the configurations we analyze are representative of the system's behavior at each studied temperature, albeit within the constraints imposed by finite simulation durations. This methodology has been shown to be effective and is widely adopted in the computational study of amorphous materials [48, 53-55].

*Determine that atoms belong to clusters:* To quantify atomic aggregation and dissociation processes, we defined direct atomic contacts based on interatomic distances, using a critical cutoff distance. This cutoff was selected as the inflection point (~2.80 Å) in the interatomic potential, which corresponds to the first peak position in the radial distribution function and is commonly adopted in similar studies of amorphous solids [34, 56-59]. According to this definition, an atom is considered part of a cluster if it maintains a chain of direct contacts connecting it continuously to all other atoms within the cluster. High temporal resolution outputs of atomic coordinates enabled precise tracking of temporal changes in cluster size due to successive dissociation and association events. This allowed for accurate calculation of corresponding rate constants within a kinetic master equation framework.

*Dynamic Mechanical Spectroscopy (MD-DMS) Simulations:* To directly probe the viscoelastic response, we carried out MD-DMS simulations. At a given temperature $T$, we apply a sinusoidal strain $\varepsilon_{xy}(t) = \varepsilon_A \sin(2\pi t/t_p)$ by simple shear method, with a period $t_p$ (related to the loading frequency $f = 1/t_p$) and a maximum value $\varepsilon_A$ of loading strain, which means the atomic affine displacement in direction $x$ increases linearly along the y direction. Then, the phase shift $\delta$ between the loading strain $\varepsilon(t)$ and the corresponding stress $\sigma(t)$ were collected for further analysis. In all the simulations, we set $\varepsilon_A = 0.75\%$ which is in the linear elastic region. For each simulation encompassing 10 loading periods, we systematically collected the relevant data. Then, the function $\sigma(t) = \sigma_0 + \sigma_A \sin(2\pi t/t_p + \delta)$ were employed to fit the mean corresponding stress. From the fitted parameters $\sigma_A$ and $\delta$, we calculated the storage modulus,

$$G' = \sigma_A/\varepsilon_A * \cos(\delta),$$

and the loss modulus

$$G'' = \sigma_A/\varepsilon_A * \sin(\delta).$$

We performed these oscillatory shear measurements at different temperatures and periods ($t_p$) for both large ($N=54,000$ atoms) and small ($N=16,000$ atoms) samples. In total, eight samples were considered: three large samples (L-50, L-100, and L-1000, corresponding to different periods $t_p$) and five small samples (S-100, S-200, S-500, S-2000, and S-4000). For instance, S-4000 refers to a small sample with $t_p = 4000$ ps. By integrating structural and viscoelastic analyses over a broad range of conditions, our methodology comprehensively captures the interplay between structural evolution (particularly cluster percolation dynamics) and the mechanical response observed upon approaching and traversing the glass transition regime.

## Acknowledgments

This research was supported by the National Natural Science Foundation of China (Grant Nos. 51971120, 51901139, U1902221) the Taishan Scholars Program ofThis research was supported by the National Natural Science Foundation of China (Grant Nos. 51971120, 51901139, U1902221) the Taishan Scholars Program of Shandong Province (tsqn201909010) and the Key Basic and Applied Research Program of Guangdong Province (2019B030302010), the Key R&D Program of Shandong Province (2022CXGC020308). A.Z. gratefully acknowledges funding from the European Union through Horizon Europe ERC Grant number: 101043968 "Multimech", from US Army Research Office through Contract No. W911NF-22-2-0256, and from the Niedersächsische Akademie der Wissenschaften zu Göttingen in the frame of the Gauss Professorship program. C.N. acknowledges support from the Royal Academy of Engineering under the Research Fellowship scheme.

**Author contributions**

A. Z., K. S. and L. H. designed the research. W. C. performed the simulations. W. C. and Z. W. analyzed the data and prepared the figures. A. Z., Z. W. and W. C. wrote the paper. C. N. provided the code and advised on the study and the manuscript. All authors discussed the data and reviewed the manuscript.

**Data, Materials and Software Availability**

The datasets generated and/or analyzed during this study are available from the corresponding author on reasonable request. The codes of the computer simulations are available from the corresponding author upon request.